\documentclass[useAMS,usenatbib,usegraphicx]{mn2e}
\usepackage{epsfig}
\usepackage{amsmath} 
\usepackage{rotating}           
\usepackage{color}     
\usepackage{graphicx}
\usepackage{times}
\usepackage{upgreek} 

\def\kms{km ${\rm s}^{-1}$}

\def\ch2{$\chi^2$}
\def\dg{$^{\circ}$}

\def\Mo{M$_\odot$}

\def\kms {\hbox{${\rm km\ s}^{-1}$}}

\def\ccm {$\hbox{{\rm cm}}^{-3}$}    
\def\scm  {$\hbox{{\rm cm}}^{-2}$}    

\def\WAT {\hbox{${\rm H_{2}O}$}} 

\def \AL {$\alpha $}     
\def \HI {H{\sc \,i}}

\def\lapp{\ifmmode\stackrel{<}{_{\sim}}\else$\stackrel{<}{_{\sim}}$\fi}
\def\gapp{\ifmmode\stackrel{>}{_{\sim}}\else$\stackrel{>}{_{\sim}}$\fi}

\title[Redshifted \HI~21-cm absorption profiles]{A comparative study of intervening and associated \HI\ 21-cm absorption profiles in redshifted galaxies}

\author[S. J. Curran,  et al.]{S. J. Curran$^{1}$\thanks{Stephen.Curran@vuw.ac.nz},  S. W. Duchesne$^{1}$, A. Divoli$^{2}$  and J. R. Allison$^{3}$\\
$^{1}$School of Chemical and Physical Sciences, Victoria University of Wellington, PO Box 600, Wellington 6140, New Zealand\\
$^{2}$Pingar, 55 Anzac Ave, Auckland 1010, New Zealand\\
$^{3}$CSIRO Astronomy and Space Science, PO Box 76, Epping NSW 1710, Australia}

\begin{document}

 \date{Accepted ---. Received ---; in original form ---}

\pagerange{\pageref{firstpage}--\pageref{lastpage}} \pubyear{2016}

\maketitle

\label{firstpage}

\begin{abstract}
  The star-forming reservoir in the distant Universe can be detected through \HI\ 21-cm absorption arising from either
  cool gas associated with a radio source or from within a galaxy intervening the sight-line to the continuum source. In
  order to test whether the nature of the absorber can be predicted from the profile shape, we have compiled and
  analysed all of the known redshifted ($z\geq0.1$) \HI\ 21-cm absorption profiles.  Although between individual spectra
  there is too much variation to assign a typical spectral profile, we confirm that associated absorption profiles are,
  on average, wider than their intervening counterparts. It is widely hypothesised that this is due to high velocity
  nuclear gas feeding the central engine, absent in the more quiescent intervening absorbers. Modelling the column
  density distribution of the mean associated and intervening spectra, we confirm that the additional low optical depth,
  wide dispersion component, typical of associated absorbers, arises from gas within the inner parsec. With regard to
  the potential of predicting the absorber type in the absence of optical spectroscopy, we have implemented machine
  learning techniques to the 55 associated and 43 intervening spectra, with each of the tested models giving a
  $\gapp80$\% accuracy in the prediction of the absorber type. Given the impracticability of follow-up optical
  spectroscopy of the large number of 21-cm detections expected from the next generation of large radio telescopes, this
  could provide a powerful new technique with which to determine the nature of the absorbing galaxy.
\end{abstract}

\section{Introduction} 
\label{intro}

Cool neutral hydrogen (\HI), the raw material for star formation, is traced through the absorption of radio continuum
radiation by the atoms undergoing the 21-cm spin-flip transition. The continuum can by intercepted by gas {\em
  associated} with the host galaxy of the radio source or  within a galaxy {\em intervening} the line-of-sight to a more distant
source.\footnote{Intervening absorption usually within galaxies which exhibit damped Lyman-\AL\ absorption, which occurs
  in the UV-band and is redshifted into the optical band at $z\gapp1.7$. A damped Lyman-\AL\ absorber (DLA) is defined
  as having a neutral hydrogen column density exceeding $N_{\rm HI} = 2\times10^{20}$ atoms per \scm\ and DLAs could
  account for more than 80 per-cent of the neutral gas content in the Universe \citep{phw05}.} As well as tracing the star formation
history of the Universe (e.g. \citealt{lbz+14}) back to $z=0$, observations of \HI\ give insight into the mass assembly
and distribution of galaxies (e.g. \citealt{rab+04}), a means of detecting the Epoch of Re-ionisation
(e.g. \citealt{cgfo04}), in addition to the potential to obtain highly accurate measurements of the fundamental
constants of nature at large look-back times (e.g. \citealt{cdk04}).

The detection of distant galaxies through 21-cm absorption is a science goal of the forthcoming Square Kilometre
Array (SKA, \citealt{msc+15}), which, through its large instantaneous bandwidth and large field-of-view, will avoid
observational biases introduced by the conventional requirement of an optical redshift to which to tune the receiver:
 \begin{itemize}
 \item[--] For the associated systems, the optical pre-selection biases toward high ultra-violet luminosities in the
   source rest-frame, which can be sufficient to ionise all of the neutral gas within the host galaxy
   \citep{cw12},  causing the observed paucity of associated 21-cm absorption at high redshift
   \citep{cww+08,cwm+10,cwsb12,cwt+12,caw+16,gd11,ace+12,gmmo14,akk16}.
 \item[--] For the intervening absorbers, pre-selection using optical redshift biases against optically obscured
   sight-lines (e.g. \citealt{ehl05}), as well as absorbers rich in molecular gas \citep{cwm+06,cwc+11}. These are of
   particular interest since molecular lines provide excellent probes of the physical and chemical conditions of the gas
   (e.g. \citealt{mbb+12}), as well as the potential to obtain accurate measurements of the fundamental constants from a
   single species (the OH radical), thus eliminating line-of-sight effects which could mimic an apparent change in the
   constants \citep{dar03}.
   \end{itemize}
   In order to obtain an unbiased census of the distribution and abundance of the cool neutral gas along each
   sight-line, it is therefore necessary to dispense with the optical pre-selection of targets which has dominated
   previous 21-cm absorption searches, in favour of using wide instantaneous bandwidths free of frequency interference
   (RFI). Coverage of the whole redshift space to $z\sim1$ is already possible with the
   Australian Square Kilometre Array Pathfinder (ASKAP, \citealt{asm+15a}), although dispensing with optical spectroscopy does present an
   obstacle in determining the nature of the absorber (see \citealt{asm+15}).
 
   In both the near-by and redshifted active galaxies, 21-cm absorption profiles are often found to be broad ($\gapp150$
   \kms, \citealt{cb95}), due either to more than one deep component
   \citep{cb95,mpb+95,cmr+98,pvtc99,topk99,vpc+00,tph+02,mot+05,moss08,meo09} or additional broad, shallow ``wings'' to
   either side of the main component \citep{mir89,mot+05,sc10a,ssm+10,mht+11,ace+12,acsr13}. This broadening of the
   additional shallow component is believed to arise from cold gas in the sub-pc, fast rotating 
   central black hole accretion disk/obscuring torus, invoked by unified schemes of active galactic nuclei (AGN,
   e.g. \citealt{ant93,up95}). The fast rotating/disturbed gas in the narrow-line region can lead to the broadening of the
    \HI\ profile in AGN (e.g. \citealt{htm08}), whereas its absence in the more quiescent intervening absorbers results in generally
   narrower and less complex profiles (e.g. \citealt{gsp+09a}). In this paper we investigate the potential of using
   these differences in the associated and intervening profiles to classify the nature of a newly identified absorption
   in the absence of complementary optical data.  Such a method could provide an invaluable tool in forthcoming spectral
   scans with the next generation of large radio telescopes.
 
\section{Differences in associated and intervening 21-cm absorption spectra}

\subsection{The sample}
\label{sec:samp}

\begin{figure}
\centering \includegraphics[angle=-90,scale=0.47]{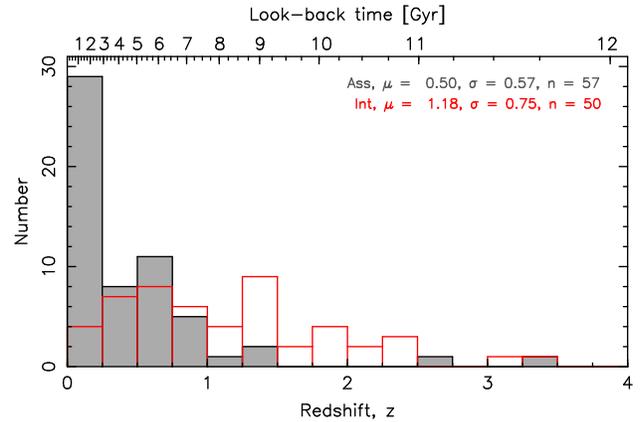}
\caption{The redshift distribution of detected associated (filled histogram) and intervening (unfilled histogram) $z\geq0.1$ \HI\ 21-cm absorbers.  The mean redshift, $\mu$, and the standard deviation, $\sigma$, for each type is shown.}
\label{distbn}
\end{figure}

As mentioned above, the use of optical redshifts can bias against the detection of 21-cm absorption and this, in
conjunction with radio flux limits, the narrow bandwidths free of RFI and the limited frequency coverage of current
telescopes, means that the detection of 21-cm absorption at $z\geq0.1$ (look-back times of $\gapp1$ Gyr) is currently
relatively rare, limited to 57 associated and 50 intervening cases (Fig.~ \ref{distbn}). 
However, we limit our analysis to these redshifts since we are interested in developing a technique to classify
absorbers for future redshifted 21-cm surveys.  Exclusion of low redshift absorption also has several advantages:
\begin{itemize}
\item[--] Minimising contamination of the sample introduced by differences between the near-by and distant absorption
  profiles.  Even if there are no intrinsic evolutionary differences, the inclusion of spatially resolved
  absorption/emission, giving several sight-lines, is expected to confuse the analysis of the unresolved sources of
  interest.
\item[--]  Dilution of the absorption profile by 21-cm emission, which is very faint beyond these redshifts (e.g. \citealt{cc15}). 
  This effect will be minimal where the spatial resolution is finest,  although, again, the mixture of unresolved and resolved features could add confusion.
\item [--] Due, at least in part, to the optical pre-selection of 21-cm absorption searches, near-by intervening
  absorption is limited to 90 sight-lines (see \citealt{cras16} and reference therein), whereas the near-by associated
  absorbers have not been compiled.  Limiting the analysis to $z\gapp0.1$ ensures approximately equal numbers of
  associated and intervening absorbers, which is important for class recognition by machine learning models (see
  Sect. \ref{mlr}).
\end{itemize}

\subsection{The spectra}

Unlike at lower redshifts, most of the $z\geq0.1$ detections are already compiled (in Tables \ref{ass_fits} \&
\ref{int_fits}, which are updated from \citealt {cw10} and \citealt{cur09a}, respectively). However, the raw data were
generally unavailable and so the spectra were acquired from the literature by digitising the available figures.
For this we used the {\sc GetData Graph Digitizer}\footnote{http://www.getdata-graph-digitizer.com/} package for all the
spectra, except those in \citet{sgmv15,ysd+16}, which were constructed from Gaussian parameters
presented.  This process was successful for 55 associated and 43 intervening spectra, with unsuccessful acquisition
resulting from noisy data or  low-quality figures. The axes were then normalised by
converting the ordinate to observed optical depth (Sect. \ref{rams}) and the abscissa to velocity dispersion, 
which was defined relative to the optical depth weighted mean velocity of the absorption
profile, $v_{\rm wm}$.  For a spectrum sampled over $i$ components this is 
\begin{equation}
v_{\rm wm} = \dfrac{\sum_{ i} \tau_{ i} v_{ i}}{\sum_{ i} \tau_{ i}} ,
\label{equ:weight}
\end{equation}
which is used to shift the abscissa so that $v_{\rm wm} = 0$.  

\begin{figure}
\centering \includegraphics[angle=0,scale=0.46]{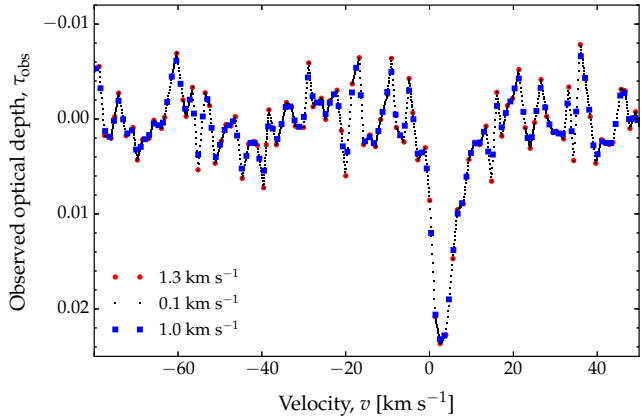}
\caption{An example of a raw spectrum (with a spectral resolution of 1.3 \kms, shown as circles) interpolated to 0.1 \kms\ (small points) and re-sampled to 1 \kms\ (square markers), used for the averaging (Sect. \ref{rams}).}
\label{spectra2}
\end{figure}

In order to allow the spectra to be inter-compared and averaged, each was oversampled and then re-binned to a common
spectral resolution (see Fig. \ref{spectra2}).
Since the spectra were shown over a variety of velocity ranges, generally proportional to the line-width, in order to
ensure that the full mean velocity range was evenly weighted, we added the typical noise level of $\tau_{\rm obs}\sim10^{-4}$ to 
each end of each spectrum to give each the same velocity range.

Half of the original spectra were presented as single or multiple Gaussian fits and so, in order to include these,
Gaussian fits were applied to all of the over-sampled spectra. Representing the spectra by Gaussians was also useful in
parameterising the spectra for comparison through machine learning (Sect. \ref{sec:fs}).  Generally, the number of
Gaussian components quoted by the authors was used and when this was not given we used the minimum number necessary for
a fit. Fitting was done using the {\sc Fityk 0.9.8}\footnote{http://fityk.nieto.pl/} package which utilises the
Levenberg-Marquardt algorithm, a least-squares fitting routine developed in particular for non-linear fitting.

\begin{figure*}
\centering \includegraphics[angle=0,scale=0.45]{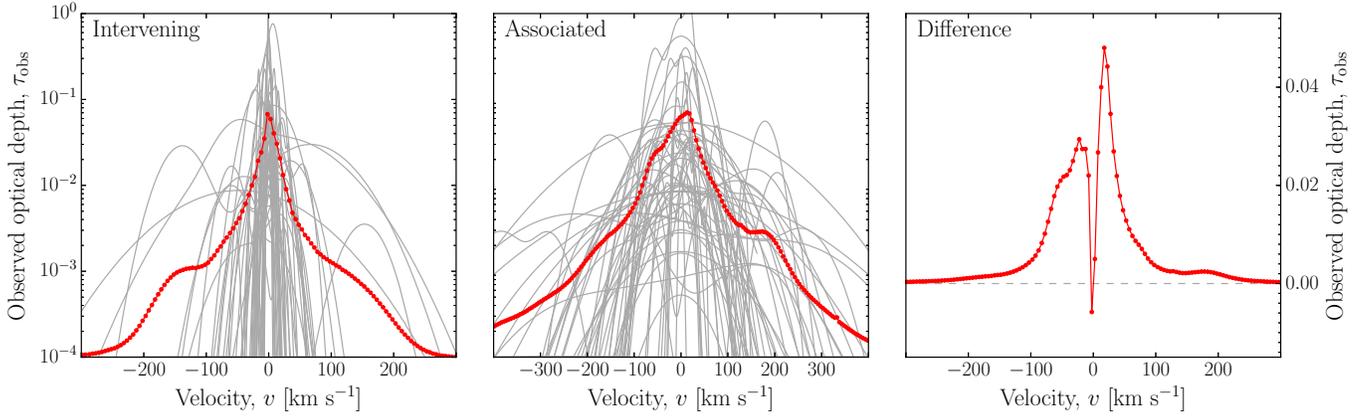}
\caption{The intervening (left), associated (middle) spectra and the mean associated minus the mean intervening spectrum
  (right, demonstrating the additional ``wings'').  The thin traces show the individual spectra and the thick traces the
  averages of these.}
\label{weight}
\end{figure*}

\begin{figure}
\centering \includegraphics[angle=-90,scale=0.48]{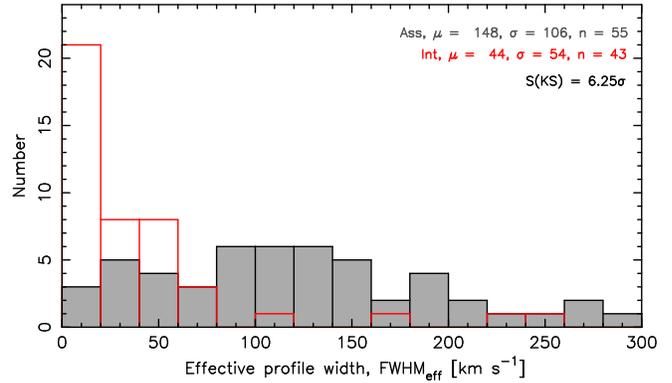}
\caption{The effective profile width, for the associated (filled histogram) and the intervening (unfilled histogram) absorbers. This is defined by  
FWHM$_{\rm eff}= \int\!\tau dv/\tau_{\rm peak}$ (e.g. \citealt{dic82,acsr13}), since for complex profiles (multiple or non-Gaussian), the
  full-width half maxima are difficult to measure in a consistent manner.}
\label{fwhm}
\end{figure}

In Fig. \ref{weight} we show the digitised spectra\footnote{These are presented individually in \citet{duc15a} and 
machine readable versions of these will be made available in a forthcoming on-line catalogue of the properties of the known redshifted \HI\ 21-cm absorbers.},
from which we see a large variation in profile shapes, although the intervening tend to have narrower velocity dispersions. 
Highlighting this, in Fig.~ \ref{fwhm} we show the distribution of the effective profile widths of the individual spectra,
from which,  we see that the associated are, on average, three times wider than the intervening profiles.
A Kolmogorov-Smirnov test gives a probability of $P({\rm KS}) = 4.17\times10^{-10}$ that the associated and intervening velocity distributions are drawn
from the same population, which is significant at $S({\rm KS}) = 6.25\sigma$, assuming Gaussian statistics. 

It is therefore clear that the profile width is related to the absorption type: The narrower profile widths of the
intervening absorbers being possibly due to the absence of the high dispersion component, in
addition to the possibility that intervening absorption may be more favourably detected in galactic disks of high
inclination (close to face-on, see \citealt{cras16}). However, the associated absorbers are seen to be roughly equally
distributed over the whole range of widths, which makes it difficult to predict the nature of the absorption based upon
the FWHM distribution alone.

\subsection{Mean spectral properties} 
\label{rams}

Although the profile widths of the individual spectra are too varied to effectively predict the absorber type, we can
average the spectra in order to explore any strong statistical difference between the types. Since the peak
optical depth is not necessarily the centroid of the absorption, we average the spectra by the mean weighted absorption,
where the zero velocity offset is defined by the median of the velocity integrated optical depth (Equ. \ref{equ:weight}), giving a more
consistent measure of redshift (e.g. \citealt{tmw+06}).
 
\begin{figure*}
\centering \includegraphics[angle=-90,scale=0.67]{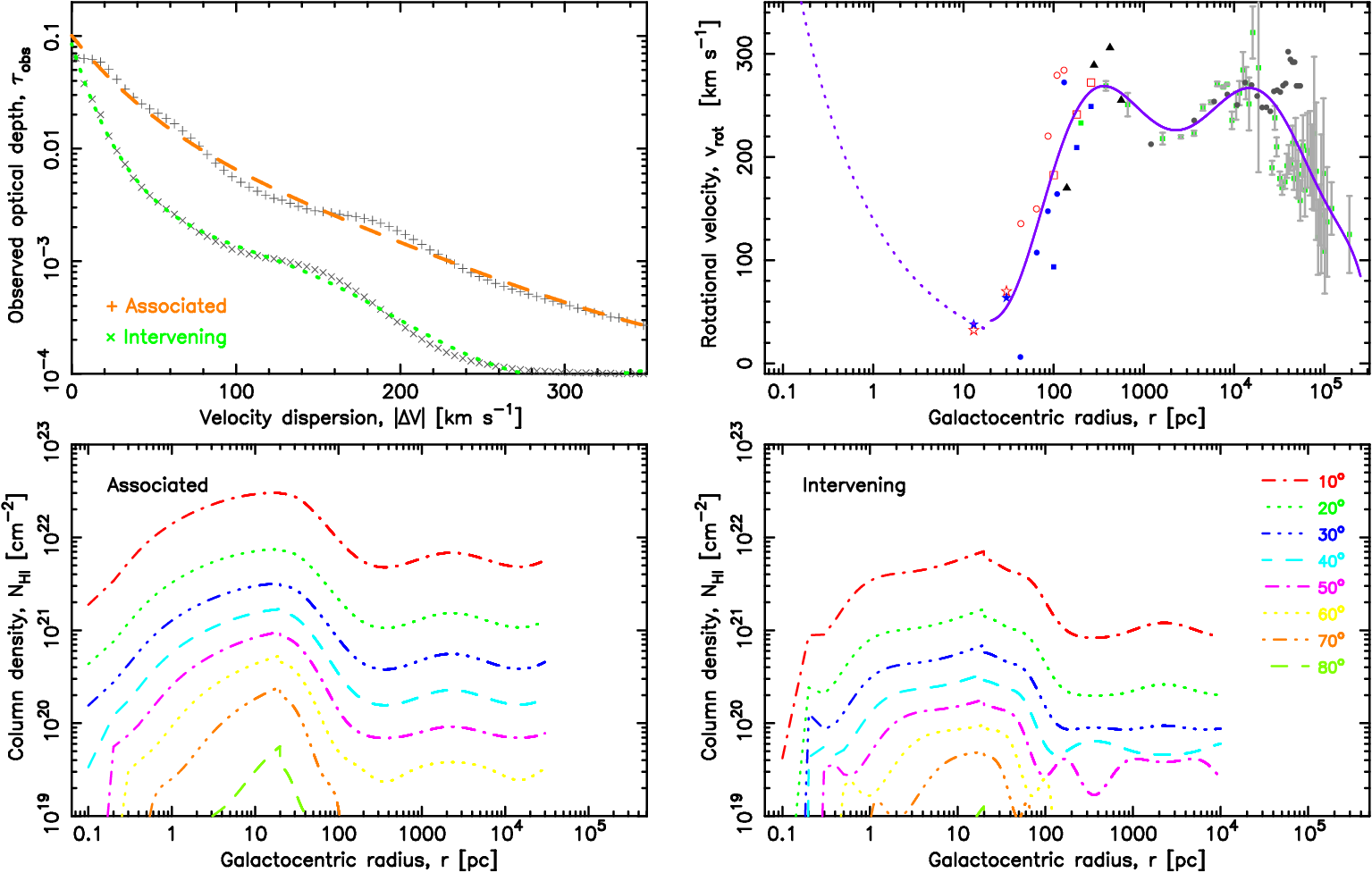}
\caption{Top left: The mean associated and intervening spectra (where the receding and approaching data are
  averaged), overlain with low order polynomial fits. Top Right: The
  rotation curve of a galaxy, based upon data from the Milky Way (shown with error bars, \citealt{bck14})
  and the Circinus galaxy (compiled from 
  \citealt{osmm94,cjrb98,dfr+98,mktg98}, see \citealt{ckb08}).  The full curve shows a polynomial fit to the large-scale
  Milky Way and Circinus data and the broken curve shows the scaled Keplerian orbit of the \WAT\ masers within the
  central 0.4 pc of Circinus \citep{gbe+03}, extrapolated to 20 pc to provide continuity.  Bottom left: The
  derived column density distribution at various disk inclinations for the associated absorbers. Bottom right: For the
  intervening absorbers. 
}
\label{4-smooth}
\end{figure*}

From the mean spectra (shown by the thick traces in Fig. \ref{weight}), it is clear the associated spectrum has
additional low optical depth gas at large velocity dispersions ($|\Delta V| \gapp200$ \kms).  As discussed in
Sect. \ref{intro}, it has been hypothesised that this is due to additional fast moving neutral gas close to the nucleus,
probably associated with the obscuring torus/accretion disk.  In order to model this additional absorption, as a
starting point in Fig.~\ref{4-smooth} (top left) we show the mean profiles, where the blue and redshifted components
have been averaged.  We then convert each optical depth--dispersion distribution to a column density--radial
distribution, by quantifying how the velocity of the gas is expected to vary with galactocentric radius.  In
Fig.~\ref{4-smooth} (top right)
we fit a polynomial to the velocity distribution of the Milky Way \citep{bck14}. Being a large spiral this may not
represent an accurate depiction of the large-scale rotation curve of an AGN host. However,
although early-type galaxies exhibit a variety of rotation curves, many exhibit similar curves to that of the Milky Way \citep{nvs+07}. 
 Furthermore, since we are interested in comparing the associated and intervening absorbers (which themselves 
may arise in a wide variety of galaxy types, e.g. \citealt{wtsc86,mmv97,pw97,hsr98,jbm99}), the Milky Way provides a
classic example of the rapid velocity increase within the central $\lapp100$ pc, before reaching velocities of $200 - 300$ \kms\  at $\gapp100$ pc \citep{nvs+07}.

Since the rotation curve of the Milky Way is only well mapped beyond $r\gapp200$ pc, we supplement this with data from
the Circinus galaxy, a near-by spiral in which the rotation curve at inner radii is readily available.  Circinus
is known to host a Seyfert\,2 nucleus \citep{mo90,osmm94} and so may at least provide a reasonable model of the inner
regions of the associated absorbers.  In order to match the velocities between the two galaxies, each of the Circinus
values have been upscaled by a factor of 1.7, which is close to the value expected based upon the scaling ratio between
the nuclear black hole and host galaxy mass (e.g. \citealt{fm00,bta+15}).\footnote{For the Milky Way $M_{\rm BH}
  \approx4\times10^6$\,\Mo\ (e.g. \citealt{rb04}), cf.  $1.7\times10^6$\,\Mo\ for Circinus \citep{gbe+03}. Assuming
  circular orbits, this gives $v_{_{\rm MW}}/v_{_{\rm Circinus}} \approx \sqrt{4/1.7} \approx 1.5$.}

Once the velocity is mapped, the column density is obtained from
\begin{equation}
N_{\text \HI}  =1.823\times10^{18}\,T_{\rm  spin}\int\!\tau\,dv,
\label{enew_full}
\end{equation}
where $T_{\rm spin}$ is the spin temperature of the gas, which is a measure of the excitation from the lower hyperfine
level \citep{pf56,fie59}, and $\int\!\tau dv$ is the velocity integrated optical depth of the absorption. The observed
optical depth is related to this via
\begin{equation}
\tau \equiv-\ln\left(1-\frac{\tau_{\rm obs}}{f}\right) \approx  \frac{\tau_{\rm obs}}{f}, {\rm ~for~}  \tau_{\rm obs}\equiv\frac{\Delta S}{S_{\rm obs}}\lapp0.3,
\label{tau_obs}
\end{equation}
where the covering factor, $f$, is a measure of the fraction of observed background flux ($S_{\rm obs}$) intercepted by
the absorber.  In the optically thin regime (where $\tau_{\rm obs}\lapp0.3$), Equ. \ref{enew_full} can be rewritten as
$N_{\text \HI} \approx 1.06 \times 1.823\times10^{18}\,(T_{\rm spin}/f) \tau_{\rm peak}\,\Delta V$, where $\Delta V$ is
the dispersion of the absorption.  Assuming that the peak of the absorption occurs, on average, in the centre of the
galaxy and that the gas is dynamically coupled to the sub-kpc rotation, the dispersion is related to the rotational
velocity via $\Delta V = v_{\rm rot}\cos i$.  To obtain the column density, we assume a constant $T_{\rm spin} = 500$~K
across the disk, since this is a constant 250--400~K across the Milky Way \citep{dsg+09} and a constant $T_{\rm spin}/f
\sim1000$~K ($T_{\rm spin} \lapp1000$~K) across external galaxies \citep{cras16}, with a mean of $T_{\rm spin}/f
\approx2000$~K ($T_{\rm spin}\lapp2000$~K) at higher redshift (in damped Lyman-\AL\ absorbers).
Higher spin temperatures close the AGN, would lead to higher column densities in the associated absorbers and so those
derived (Fig.~\ref{4-smooth}, bottom left) should be considered lower limits.
For the covering factor we assume full coverage ($f=1$) for the mean face-on ($i=90$\dg) disk, so that $\tau = {\tau_{\rm obs}}/{\sin^2\!i}$ \citep{cur12}.

Assuming that the gas remains sufficiently cool and neutral to exhibit detectable 21-cm absorption within the inner
$\sim1$~pc, where the kinematics are dominated by Keplerian rotation around a massive compact object (the central
black hole), this simple model does indeed suggest that additional high dispersion gas in the associated absorbers arises from a central
component, rather than being dominated by orientation effects (Sect. \ref{sec:samp}). 

Furthermore, the high column densities for the low inclination associated absorbers are consistent with what we expect
from the Milky Way, in which the volume density of the neutral gas exhibits an exponential decrease with galactocentric
radius according to $n = n_0\,e^{-r/R}$, where $n_0 = 13.4$ \ccm\ and the scale-length $R = 3$ kpc \citep{kdkh07}.  From
$N_{\text{\HI}}\equiv\int\!ndl$, the column density has a maximum at $i=0$\dg, given by $N_{\text{\HI}}=
n_0\int_{0}^{\infty}e^{-r/R}dr = n_0R = 1.2\times10^{23}$ \scm, which remains high to large radii, since this is the
total volume density integrated over the path length.

\subsection{Redshift evolution}
\label{sec:re}

In addition to any possible differences in the intervening and associated profiles due to the presence of gas
in close proximity to the AGN,  it is also possible that differences could be introduced by redshift evolution.
Unfortunately, the associated absorbers are predominantly at $z\lapp1$, due to higher
redshifts preferentially selecting the most UV luminous sources (Sect .\ref{intro}). For the intervening absorbers, however, the
sample is split in half at $z\sim1$, which, at a look-back time of LBT\,$\approx8$~Gyr, is close to half the age of the Universe,
which allows us to compare the profiles between these two epochs.

\begin{figure}
\centering \includegraphics[angle=-90,scale=0.45]{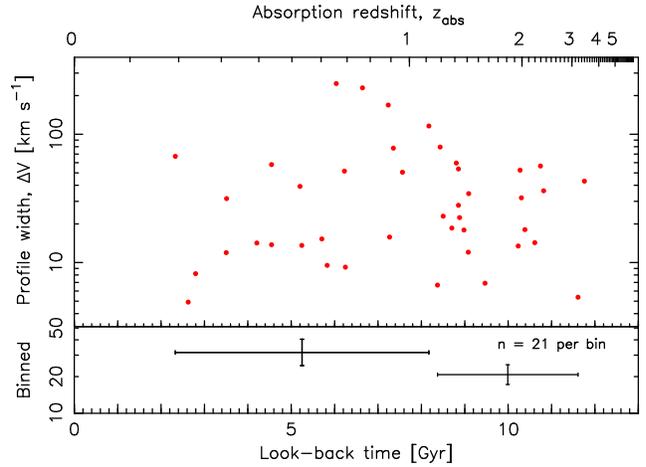}
\caption{The effective profile width versus look-back time for the intervening absorbers.}
\label{int-z}
\end{figure}

From the distribution of profile width with redshift (Fig. \ref{int-z}), 
we see no evidence of any evolution. If the thermal broadening is comparable to that introduced by the gas kinematics
(Sect. \ref{rams}), this would suggest no mean evolution in the kinetic temperature of the gas, $T_{\rm kin}$.  The
intervening absorbers are not subject to the same broadening mechanisms as the associated absorbers, where gas
kinematics, turbulence and radiative heating can be significant, leading to possible line broadening (see
Fig. \ref{fwhm}).  That is, in thermodynamic equilibrium the spin temperature is coupled to the kinetic temperature
(e.g. \citealt{lb01,rcs06}), which is given by
\[
T_{\rm kin} \leq \frac{\Delta V^2}{8\ln(2)}\frac{m_{_{\rm H}}}{k_{_{\rm B}}}~[\text{K}]\lapp 22 \Delta V^2 [\text{K for $\Delta V$ in \kms}],
\]
where $m_{_{\rm H}}$ is the mass of the hydrogen atom and $k_{_{\rm B}}$ is the Boltzmann constant. From the binned line-widths
(Fig. \ref{int-z}), we obtain $T_{\rm kin}\lapp22\,000$~K at LBTs $\lapp8$ Gyr and $T_{\rm kin}\lapp9\,000$~K at LBTs
$\gapp8$ Gyr. The limit arises due to other possible broadening mechanisms, although the absence of any increase in
profile width with redshift does not support the argument that there is an increase in temperature with redshift
(\citealt{kc02}, where the $T_{\rm spin} = T_{\rm kin} = 22 \Delta V^2$  assumption is used).  This is consistent with the argument
that the larger $T_{\rm spin}/f$ ratios measured at high redshift are dominated by lower covering factors
\citep{cmp+03}, through the geometry effects of an expanding Universe \citep{cur12}.
 
\section{Exploring Machine Learning Models for Classification}

\subsection{Motivation for using machine learning}

The main motivation for this study was to determine whether the absorber type can be predicted from the profile
properties without {\em a priori} knowledge of the emission redshift from an optical spectrum. As seen in Fig
\ref{weight}, however, the individual spectra are too varied to permit this, with the associated absorbers spanning the full range of
line-widths (Fig. \ref{fwhm}).  We therefore apply machine learning techniques with the aim of building a classifier
which can be used to make such a prediction.  While the data set is quite limited, with $\lapp50$ useful spectra in each
class, we can explore how feasible machine learning models are and the potential for prediction, particularly as more
data are added. We use the {\sc weka} package \citep{hfh+09}, a suite of machine learning algorithms designed
for data mining tasks.

\subsection{Models}
\label{mas}

A model is the result of the training data, the features selected to represent the data, the algorithm used, as well as
a number of parameters set for the algorithm.

\subsubsection{Feature Selection}
\label{sec:fs}

\begin{figure}
\centering \includegraphics[angle=0,scale=0.47]{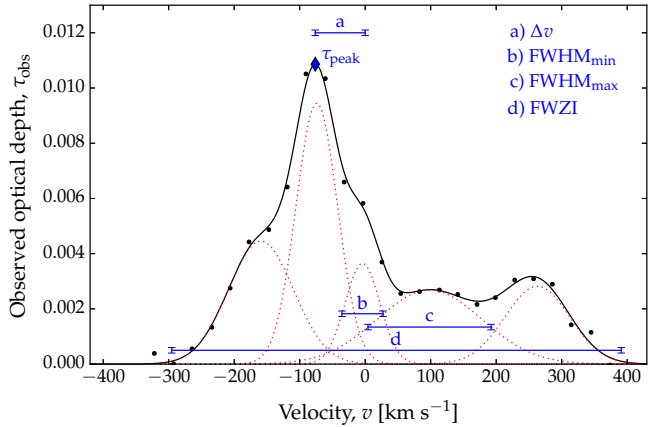} 
\caption{An example of a complex ($n_{\rm g}=5$) absorption profile, 
labelled with the features
  used for the machine learning: $\Delta v$ the offset of the component from $z_{\rm weight}$ (a), the minimum (b) and
  maximum (c) FWHM of the profile and the Full-Width at Zero Intensity (d, where $\tau_{\rm obs}<10^{-4}$).}
\label{features}
\end{figure}

A crucial step in machine learning is the selection of {\em features}. These represent each datum and therefore need to
be discriminating and informative. The feature set forms the input given to the selected algorithm and it is the
combination of the features which allow different algorithms to discriminate between classes.  Here we identified a
mixture of such features for each spectrum, illustrated in Fig. \ref{features} and listed in Tables~\ref{ass_fits} and
\ref{int_fits}.  
\begin{table*} 
\centering 
\begin{minipage}{160mm}
  \caption{The features of the $z\geq0.1$ associated 21-cm absorbers. The first column gives the IAU name, followed by
    the mean-weighted absorption redshift (see Sect. \ref{rams}) and the number of Gaussian components required to fit
    the spectrum, $n_{\rm g}$. The following columns give the Full-Width at Zero Intensity, the peak observed optical
    depths, the average offset of the components from $z_{\rm weight}$ and the full-width half maxima, respectively (see
    Fig. \ref{features}).  The last column gives the reference for the 21-cm absorption.}
\begin{tabular}{@{}ll  cc   ccc   r r rrr  l@{}} 
\hline
IAU & $z_{\rm weight}$ & $n_{\rm g}$  & FWZI    & \multicolumn{3}{c}{$\tau_{\rm peak}$} & $\overline{\Delta v}$ & $\overline{\Delta v}$/FWZI & \multicolumn{3}{c}{FWHM} [\kms] & Ref.\\
       &                           &           &             & ave & max & min                                   &         [\kms]                               &    [\kms]                            & ave & max & min & \\
\hline
B0023--26   &  0.321409   & 2  & 482 & 0.00573 & 0.00944 & 0.00201 & -39 &  -0.081 & 149 &  173 & 126 & V03\\
B0108+388  & 0.668475   & 2   &426  & 0.0423  & 0.0516   &  0.0329   & - 5 &  -0.011 & 107 & 139  & 76   &  O06\\ 
J0141+1353 & 0.620390   & 2   & 92   & 0.0155   & 0.0169   & 0.0141   &   3 & 0.031   &  21 & 35 & 8   & V03 \\
B0316+16    &  0.906947   & 3  & 243  & 0.0131   &0.0273   & 0.00226  &-7   & -0.031   & 42  & 95  &10 &  S10 \\
J0410+7656 &   0.598867 & 3   & 785  &0.00656  & 0.0141 &0.00268   & -154   & -0.20   & 80  & 104  & 61 & V03 \\
J0414+0534  &  2.636414  & 2 & 575 & 0.0131  & 0.0151     & 0.0111    &  29     & 0.051    &  147  & 154 & 141& M99\\ 
B0428+20 &0.220358  & 2 &  951 & 0.00337 & 0.00467 & 0.00207 & 69  &  0.073  & 283 & 299 & 268 & V03 \\
B0500+019  & 0.584693 & 2 & 224   & 0.0288     & 0.036     & 0.0216   & -16 & -0.071  & 54  & 63 & 45& C98\\
B0758+143   &  1.194147 & 5 & 686 & 0.00461 & 0.00944 & 0.00268 & 25    & 0.036     & 112 & 188 & 63  & I03\\
J0834+5534  & 0.240669  & 1 & 443 & 0.00281  &  0.00281 & 0.00281 & 0        &  ---         & 202 & 202 & 202  & V03 \\
B0839+458 &  0.192255 & 2 & 325    &0.178      & 0.299       & 0.0574   & -18.4 &-0.057   & 79 & 84  & 74 & G15\\
B0859+032	&  0.288116   &   2 & 547 & 0.0474 & 0.0560     & 0.0387  &14   & 0.025    & 73  & 92 & 54 & Y16 \\ 
J0901+2901  &  0.193870 & 1 & 179 & 0.00053   &  0.00053   &0.00053   & 0 & --- & 117 & 117 & 117&V03\\ 
B0902+34 & 3.396779   & 1 & 700 & 0.00849 &  0.00849 &0.00849 & 0 &---   & 277 & 277 & 277 & U91\\
J0909+4253 &  0.670341& 2 & 276 & 0.00647 & 0.0102 & 0.00273 & 55 & 0.20 & 61 & 101 & 20  & V03 \\
B20917+27B & 0.206698 &   2   &  96 & 0.0612 & 0.0819   &0.0405 &  -0.9   & -0.0089 &230  & 33 & 13 &  Y16 \\
J0942+0623  &  0.123206 &    3  & 191 & 0.723   & 0.896    & 0.381  & -13 &-0.068  &  30 & 45 & 14 & S15\\
B1003+35  &  0.099742 & 3 & 558 & 0.0139   & 0.0203 & 0.00506 & 42 & 0.075 & 117 & 203 & 65 & V89 \\
B1107--187  & 0.491705 & 2 & 76 & 0.0161    & 0.0189 & 0.0133  & 0  & 0.0046 & 19 & 22 & 17 & C11b\\
J1120+2736 & 0.111720  & 1 & 221 & 0.159 & 0.159 & 0.159 & 0 & --- & 67 & 67 & 67 & G15\\
J1124+1919  &0.165161   & 6 & 77 & 0.0407 & 0.101 & 0.00549 & 12 & 0.15 & 9 & 13 & 7 & G06 \\ 
B1147+557     &  0.138297 & 3 & 234 & 0.0217 & 0.0494 & 0.0204 & 9 & 0.038 & 34 & 42 & 22 & C11 \\ 
B1126+569   &  0.891604  & 2  & 464 & 0.0170 & 0.0276 & 0.0063 & -25 & -0.053 & 137 & 170 & 104 &   Y16 \\ 
B1142+052    &  1.343073 & 3 & 157 & 0.00524 & 0.00707 & 0.00352 &  6 & 0.037 &31 & 39 & 15 & K09 \\
J1202+1637 &  0.118568 & 2 & 622 & 0.0213 & 0.0279 & 0.0147 & 6 &  0.0091 &  179 & 231 & 127& G15\\
B1203+645    & 0.371883 &    1 & 405 & 0.00385 &  0.00385 & 0.00385 & 0 & --- & 177  & 177 & 177 & V03 \\
B1206+469    & --- &     --&   \multicolumn{9}{c}{\sc could not be fit/spectrum of too poor quality}  & G15\\
B1244+49	&0.205956  & 4 & 686 & 0.000591 & 0.000657 & 0.000463 & 22 &  0.032 & 192 & 260 & 119 & G15\\
J1254+1856	& ---& -- &  \multicolumn{9}{c}{\sc could not be fit/spectrum of too poor quality} & G15\\
J1301+4634 & 0.205041 & 3 &901 & 0.00777 & 0.0108 & 0.00465 & 7 & 0.0081 & 198 & 298 & 111 & G15\\
J1326+3154  & 0.368430 & 1 &  459 & 0.00170 & 0.00170 &0.00170 & 0 & --- &  227 & 227 & 227&  V03\\
J1347+1217 &  0.121924 & 3 & 336  &  0.00493 & 0.0107 & 0.0018 & 58 & 0.150 & 97 & 130 & 23  & G06\\ 
B1355+441 &  0.645449 & 1 & 1085  & 0.0537 & 0.0537 & 0.0537 & 0 &--- &  359 & 359 & 359 & V03\\
J1357+0046 &  0.796663 & 2 & 295    &  0.0109 & 0.013 & 0.0088 & 25  & 0.085 & 76 & 80 & 71 &   Y16\\ 
J1400+6210 & 0.430137  & 1 &411 & 0.00611 & 0.00611 & 0.00611 &0 & --- &  169 & 169 & 169 &  V03\\
J1409+3604 &  0.148418& 2 & 249 & 0.034   & 0.0373 & 0.0317 & -10 & -0.038 & 57  & 66 & 50 &  C11\\
B1413+135 &  0.246079&  2 & 72  & 0.0288   & 0.036  & 0.0216 & -16 & -0.22     & 54 & 63 & 45 & C92\\ 
J1422+2105 & 0.190425 & 1 & 544 & 0.0434 &  0.0434 & 0.0434 & 0 &--- & 183 & 183 & 183 &  G15\\
B1504+37      &  0.672634 & 3  & 258 & 0.205 & 0.335 & 0.0759 & -8 & -0.031 & 45 & 85 & 17 & C98\\
B1543+480    &  1.277005  & 2 &532 & 0.0425 & 0.0533 & 0.0317 & -20 &-0.038 & 109 & 129 & 89 &   C13 \\
B1549--79       &  1.518581 &  3 & 330 & 0.00937 & 0.0174 & 0.00439 & 12 & 0.035 & 72 & 140 & 34  & M01 \\
B1601+5252&  0.105545  &   1 & 243 & 0.00974 & 0.00974 & 0.00974 & 0 & --- & 95 & 95 & 95 & C11\\ 
B1603+609  &    0.559129    &   1 & 21 & 0.0142 &0.0142 & 0.0142 & 0 & --- &  8 & 8 & 8 &   Y16 \\ 
B1614+26      &  0.755466    &    1 &  1129 & 0.0084 & 0.0084 & 0.0084 & 0 & --- & 447 & 447 & 447 &   Y16 \\
B1649--062  & 0.236387 & 1 & 503 & 0.0238 & 0.0238 & 0.0238 &  0 &--- & 179 & 179 & 179 &  C11c\\
B1717+547   &  0.147402 & 2 & 351 & 0.0276 & 0.0326 & 0.0225 & -7 & -0.020 & 100 & 149 & 51 & C11\\  
B1814+34     &  0.243994&  1& 231 & 0.0396  &  0.0396  &  0.0396  &  0 & --- & 79 & 79 & 79& P00\\
J1815+6127 & 0.596812 & 1 & 327 & 0.0207 & 0.0207 & 0.0207 &   0 & --- &118 & 118 & 118 &  V03 \\
J1821+3942   & 0.795323  & 2 &259 & 0.00884 & 0.01 & 0.00768 & -2 & -0.0077 & 53 & 61 & 44&  V03\\
J1944+5448 & 0.258259 & 1 & 795 & 0.00864 & 0.00864 &  0.00864 &  0 & --- & 313 & 313 & 313 & V03 \\
J1945+7055   &  0.100209  & 1 & 306 & 0.55 & 0.55 & 0.55 & 0 &--- & 87 & 87 & 87 &  P99\\
B2050+36 & 0.354647& 2 & 109 & 0.125 & 0.204 & 0.0462 & -6 & -0.055 & 24 & 32 & 16& V03\\
B2121+248   & 0.107209 & 3  & 527 & 0.00217 & 0.00432 & 0.000999 & 1 & 0.002 & 124 & 200 & 70 & M89\\
B2252--089  &0.607037  & 2 &286 & 0.132 & 0.181 & 0.0839 & 0.028 8 & & 55 & 92 & 18 &  C11b\\
J2255+1313  & 0.543101 & 1 & 281 & 0.00162 & 0.00162 & 0.00162 & 0 &--- & 140 & 140 & 140 &  V03 \\
J2316+0405   & 0.219135   & 1 & 289 & 0.00305 & 0.00305 & 0.00305 & 0 &  --- & 130 & 130 & 130 & V03 \\
J2355+4950 &  0.237905 & 2 & 272 & 0.0145 & 0.0176 & 0.0114 & 59 & 0.22 & 47  & 81 & 13 & V03\\
\hline
\label{ass_fits}
\end{tabular}
{References: M89 -- \citet{mir89}, V89 -- \citet{vke+89}, U91 -- \citet{ubc91}, C92 -- \citet{cps92}, C98 -- \citet{cmr+98}, M99 - \citet{mcm98}, P99 -- \citet{ptc99}, P00 -- \citet{ptf+00}, M01 -- \citet{mot+01}, I03 -- \citet{ida03}, V03 -- \citet{vpt+03}, G06 -- \citet{gss+06}, O06  -- \citet{omd06}, K09 -- \citet{kpec09}, S10 -- \citet{ssm+10}, C11 -- \citet{css11},  C11b -- \citet{cwm+10}, C11c -- \citet{cwwa11},  C13 -- \citet{cwt+12}, G15 -- \citet{gmmo14}, S15 -- \citet{sgmv15}, Y16 -- \citet{ysd+16}.}
\end{minipage}
\end{table*} 
\begin{table*} 
\centering 
\begin{minipage}{160mm}
\caption{As Table \ref{ass_fits} but for the $z\geq0.1$ intervening 21-cm absorbers.} 
 \begin{tabular}{@{}ll  cc   ccc   r r rrr  l@{}} 
\hline
IAU & $z_{\rm weight}$ & $n_{\rm g}$  & FWZI    & \multicolumn{3}{c}{$\tau_{\rm peak}$} & $\overline{\Delta v}$ &$\overline{\Delta v}$/FWZI & \multicolumn{3}{c}{FWHM} [\kms] & Ref.\\
       &                           &           &             & ave & max & min                                   &         [\kms]                               &    [\kms]                            & ave & max & min & \\
\hline
J0108--0037   &1.370985  & 1 & 52 & 0.0731 & 0.0731 & 0.0731 & 0 & --- & 17 & 17 & 17 & G09b\\
B0132--097 & 0.764436 & 2 & 519 & 0.0294 & 0.0299 & 0.0288 & 0 &  -0.069 &  110  & 145& 74  & K03a\\
B0201+113   &  3.386789 & 2 & 143 & 0.0129 & 0.017 & 0.0088 & -2 & -0.015 & 29 &  37 & 22 & K14a\\
B0218+35      & 0.684651 & 1  & 143 & 0.0454 & 0.0454 & 0.0454 & 0 &--- & 49 & 49 & 49 &   C93\\
B0235+164   &  ---     &  --& \multicolumn{9}{c}{\sc could not be fit/spectrum of too poor quality} & R76\\
B0237--233     & ---& --  & \multicolumn{9}{c}{\sc could not be fit/spectrum of too poor quality} & K09\\
 B0248+430  & 0.394153 & 3 & 49 & 0.175 & 0.222 & 0.126 & 0 & 0.0041 &  6 &  7 & 4 & L01\\
B0311+430    &  2.289521    & 2 & 145 & 0.0101 & 0.0146 & 0.0055 & -14 & -0.099 & 37 & 40 & 34 &   Y07\\
J0414+0534  & 0.959790& 3  & 127 & 0.0113 & 0.0174 & 0.00348 & 13 & 0.11 &  26 & 31 & 19 & C07a \\ 
B0438--436  &  2.347469 & 2 & 83 & 0.00452 & 0.0059 & 0.00314 & 7  & 0.087 & 21 & 24 & 18 &  K14a\\
B0454--234 &  0.891324& 1 & 39 & 0.0130 &  0.0130 &0.0130 &  0 & --- &15 & 15 &15  &  G12\\ 
B0458--020  &  1.560516& 1 & 17 & 0.0233 & 0.0233 & 0.0233 &  0 &--- &  7 & 7 & 7 & K09\\
....                     &  2.039484 & 2 & 54 & 0.237 & 0.334 & 0.139 & 4 & 0.069 &  12 & 13 & 10 & K14a\\
B0738+313   &0.220999   & 2 & 14 & 0.0475 & 0.0634 & 0.0316 &  0 & 0.0071 & 4 & 5 & 3 &   K01b \\
J0804+3012  &1.190890   & 2 & 198 & 0.00297 & 0.00392 & 0.00201 & -9 & -0.045 & 66 & 89 & 43 & G09b\\
J0808+4950   & 1.407309 & 1 & 28 & 0.00766 &  0.00766 &0.00766 & 0 & --- & 11 & 11 & 11 & G09b\\
B0809+483  & 0.436899 &2  & 347 & 0.0101 & 0.0167 & 0.00354 &47 &  0.13  & 56 & 72 & 39 & B01 \\ 
B0827+243  & 0.524762    & 1  & 89 & 0.00578 & 0.00578 &  0.00578 &   0 & --- & 37 & 37 & 37 & K01\\
J0849+5108   & 0.311991  & 2 &  43 & 0.0415 & 0.0602 & 0.0302 & -5 &-0.11 &  9 & 12 & 5 &  G13\\  
J0850+5159 &  1.326818   & 3 & 150 & 0.274 & 0.441 & 0.126 &  -1 &-0.0053 & 28 & 47 & 13 & G09b\\
J0852+3435  & 1.309508 & 2 & 193 & 0.0868 & 0.0885 & 0.0851 & -8 & -0.040 & 42 & 62 & 22 &  G09b\\
B0927+469  & 0.621550 & 1   & 22 & 0.0323 & 0.0323 & 0.0323 & 0 &  --- & 9 & 9& 9 & Z15\\
B0952+179  & 0.237808  & 1 & 19 & 0.013 & 0.013 &  0.013 &  0 & --- & 8 & 8 & 8 & K01a\\
B1055+499 &    1.211757 & 2  &69 & 0.0121 & 0.0188 & 0.00538 & -4 & -0.059 & 21 & 25 & 16 &  G09b\\
B1127--145    &  0.313012  & 7 & 123 & 0.0522 & 0.127 & 0.00327 &  1 & 0.0070 &111 & 15 & 6 & C00\\
B1157+014 & 1.943628 & 2 & 33 & 0.0564 & 0.0698 & 0.0429 & 2 &  0.053 &  8 & 8 & 7 & K14a\\
B1229--0207  & --- & -- & \multicolumn{9}{c}{\sc  could not be fit/spectrum of too poor quality} & L01\\
B1243--072 &  0.437217 & 1 & 39 & 0.0684  &0.0684  & 0.0684  &0 &  --- & 13 & 13 & 13 & L01\\
B1252+4427  & 0.911272 & 3 &229 & 0.0183 & 0.0277 & 0.00995 & 2 & 0.0083 & 41 & 61 & 27 & G12\\
B1328+307 &  0.692150 & 1 & 8 & 0.093 &  0.093 & 0.093 &  0   & --- & 9 & 9& 9 &  D78\\
B1331+17 &  --- & -- &  \multicolumn{9}{c}{\sc  could not be fit/spectrum of too poor quality} & B83\\
J1337+3152 &   3.17447   &   1 & 17 & 0.611 & 0.611 &  0.611 &  0 &--- & 5 & 5 & 5 &  S12\\  
B1406-076 & 1.274564  & 1 & 71 & 0.0160 & 0.0303 & 0.00601 & -1 &-0.0089  & 11  & 12 & 10   & G12\\
B1430--178 &  1.326455  & 2 & 119 & 0.00158 & 0.00215 & 0.00101 & -6   & -0.054 & 37 & 57 & 16 & K09\\
J1431+3952  &0.601849  & 2 & 39 & 0.226 & 0.227 & 0.224 & 1 & 0.026 & 7 & 8 & 6 & E12 \\
J1443+0214  &  0.371540  & 2 & 42 & 0.256 & 0.402 & 0.109 &  0 & -0.0060 &10 & 14 & 6 &  G13\\
B1621+047   &  1.335695& 2 & 65 & 0.0322 & 0.0413 & 0.0231 & 5 &  0.068 &  15 & 17 & 12 &  G09a\\
B1622+238  &  0.655943   & 1 & 500 & 0.00877 & 0.00877 & 0.00877 & 0 &--- & 233 &233 & 23 &  C07b\\
B1629+120  &  0.531764 & 2 & 44 & 0.0211 & 0.0261 & 0.0161 & 1 & 0.027 &  13 & 21 & 6  & K03b\\
B1755+758   &  1.970875 & 1 & 137 & 0.0227 & 0.0227 &0.0227 &  0  & --- &49&49&49 &  K14b\\
B1830--21 &0.885469 & 2 & 574 & 0.0338 & 0.0368 & 0.0307 & -18 & -0.032 & 132 & 197 & 67 & C99\\
...                  &  0.192504 & 3 & 124 & 0.0116 & 0.0134 & 0.00954 & 2 & 0.017 & 24 & 28 & 15 & L96\\
B1850+402 & 1.989599 & 2 & 88 & 0.0784 & 0.0956 & 0.0612 & 5 & 0.058 & 18  & 20 & 16  & K14b\\
B2003--025 & 1.410732 & 2 & 64 & 0.00498  & 0.00587 & 0.00408 & -5 &-0.082 & 18 & 24 & 12 & K09\\
B2029+121 & 1.115735  & 2 & 266 & 0.0156 & 0.0186 & 0.0125 & -3 & -0.012 & 70 & 78 & 61.4 & G12\\
B2039+187 &  2.191798  & 1  & 38 & 0.0320 &  0.0320 & 0.0320 &0 & --- & 13 & 13 & 13 &    K12 \\
J2340--0053 & 1.360890& 3 &6 & 0.400      & 0.82 & 0.182 & 0 &  0.022 & 3 & 6 & 1  & G09b\\
B2351+456  & ---   & -- & \multicolumn{9}{c}{\sc could not be fit/spectrum of too poor quality} & D04\\
B2355--106 &1.173038  &  1 & 17 & 0.0333  & 0.0333  &  0.0333  &  0  &--- & 6 & 6 & 6  &  G09b\\
\hline
\label{int_fits}
\end{tabular}
{References: R76 -- \citet{rbb+76}, D78 -- \citet{dm78}, B83 -- \citet{bw83}, C93 -- \citet{cry93}, L96 -- \citet{lrj+96}, C99 -- \citet{cdn99}, B01 -- \citet{bdv01}, L01 -- \citet{lb01}, K01a -- \citet{kc01},  K01b -- \citet{kgc01},  K03a -- \citet{kb03}, K03b -- \citet{kc02}, D04 -- \citet{dgh+04}, C07a -- \citet{cdbw07},  C07b -- \citet{ctp+07},   Y07 -- \citet{ykep07}, G09a -- \citet{gsp+09},  G09b -- \citet{gsp+09a},  K09 -- \citet{kpec09},  E12 -- \citet{ekp+12}, G12 -- \citet{gsp+12}, K12 -- \citet{kem+12}, S12 -- \citet{sgp+12},  G13 -- \citet{gsn+13}, K14a -- \citet{kps+14},  K14b -- \citet{kan14}, Z15 -- \citet{zlp+15}.}
\end{minipage}
\end{table*} 
 
Although each spectrum can be quantified through standard parameters such as the number of Gaussians ($n_{\rm g}$), the
velocity offset ($\Delta v$), the peak optical depth ($\tau_{\rm peak}$) and the FWHM, this will present an issue for
the machine learning in that for $n_{\rm g}>1$ there will be several parameters classified as $\Delta v$, $\tau_{\rm
  peak}$ and the FWHM, which must be independently flagged as such while also being compared to other spectra.  For example,
if both spectra {\sf A} and {\sf B} each have two components, then the first component of {\sf A}, FWHM$^{\sf A}_{1}$,
must be compared with both FWHM$^{\sf B}_{1}$ and FWHM$^{\sf B}_{2}$, as must FWHM$^{\sf A}_{2}$, while all retaining
their identity as representing the line-width of an unspecified component. In machine learning it is common practice to
compare global properties, which in this instance would be the total or average
FWHM.
The chosen features represent unique (or a combination of unique) properties of the spectrum, with no
prior expectations of how they will perform. Therefore, in addition to features analogous to those standard
in quantifying the spectral properties, such as the line-width and optical depth, we include the average offset of the
components from $z_{\rm weight}$.  This is included solely for the reason in that it is an additional property which can
be extracted, although there is the possibility that this could be non-zero for the associated spectra, where there is
additional in or outflowing gas.

In the final models we excluded $z_{\rm weight}$ itself, although it may be a powerful feature (see Fig. \ref{distbn}),
and indeed, upon testing, can increase the precision of the models (see Sect. \ref{mlr}). However, we
believe that the tendency for the associated absorbers to be detected at $z\lapp1$ is due to their optical
pre-selection, where at higher redshifts the faint optical light observed is intense ultra-violet in the
rest-frame of the source, ionsing the gas in the high redshift objects \citep{cw12}. Since, the SKA
and its pathfinders will not be reliant upon optical redshifts, through full redshift spectral scans (Sect.~\ref{intro}), 
we expect higher redshift detections to be forthcoming giving a $z_{\rm weight}$ distribution more similar to that of the
intervening absorbers.

\subsubsection{Comparison of predictive power of different features}

{\sc weka} also offers the ability to explore the predictive power of the different features via the {\em Select
Attributes} function. This is useful in removing non-contributing features, particularly for large data sets where
computational power is an issue. This does not apply to our small sample, although it is of interest to find which
features contribute most to the predictive power of the models. 
\begin{table} 
\begin{minipage}{65mm}
\caption{The rankings of the features for the whole sample and excluding the optically thick absorbers 
(in descending order for the whole sample).}
\begin{tabular}{l c c}
\hline
Feature  & \multicolumn{2}{c}{Pearson's correlation} \\
              & Whole   & Excluding $\tau_{\rm peak}\geq0.3$ \\
\hline
FWZI & 0.5552 & 0.5423\\
FWHM max & 0.5447 & 0.5367\\
FWHM ave & 0.5017 & 0.4943\\
FWHM min & 0.4444 & 0.4384 \\
$\tau_{\rm peak}$ max & 0.0907 & 0.0321 \\
$\tau_{\rm peak}$ min & 0.0904 & 0.1580 \\
$\tau_{\rm peak}$ ave  &  0.0782 & 0.0900 \\
$\overline{\Delta v}$  & 0.0540 & 0.0654 \\
$n_{\rm g}$  & 0.0420 & 0.0494 \\
$\overline{\Delta v}$/FWZI  & 0.0336 & 0.0648 \\
\hline
\label{att}
\end{tabular}
\end{minipage}
\end{table}
In Table \ref{att}, we show the rankings returned by the attribute evaluator. 
From this, we see that all features contribute to the classification. Those related to the profile
width contribute the most, while other features make relatively little contribution, including, surprisingly, the number
of Gaussians.

\subsubsection{Training Data}

Since the data set is too small to split into training {\em and} testing sets,
we use all data, 55 associated and 43 intervening spectra, for training.
The models were run for both the whole sample and with the
exclusion of spectra exhibiting optically thick components (Equ.~\ref{tau_obs}).

\subsubsection{Algorithms}

Since there is no single algorithm that is more suited to all cases
(e.g. \citealt{wol96}), nor any previous application of machine learning
to 21-cm absorption profiles in the literature, we experimented
with different algorithms in {\sc weka}. As this is a small dataset, we
kept the default parameters for each algorithm. For the same reason,
we considered all features with equal weight. Several algorithms
performed comparably well, here we report five of which
are classifiers for categorical prediction, as opposed to classifiers
for numeric prediction \citep{wf11}:

\begin{enumerate}
\item {\em Bayesian Network} from the ``Bayes'' group \citep{bou04}.  This is a probabilistic model, which through the
  learning of Bayesian nets, represents a set of random variables which may be observable quantities, latent variables,
  unknown parameters or hypotheses \citep{bou08a}.

\item {\em Sequential Minimal Optimisation} from the ``functions'' group. This algorithm solves the quadratic
  programming problem, arising from the training of support vector machines \citep{pla99}.

\item {\em Classification Via Regression} from the ``meta'' group. This algorithm
uses regression methods, where one regression model is built for each class value \citep{fwi+98}.

\item {\em Logistic Model Tree } from the ``trees'' group. This combines logistic regression and decision
  tree learning, based upon an earlier version of the tree.  Each ``leaf'' in the tree represents a model and
  the logistic variant produces a regression model at every node in the tree, which is then split \citep{lhf05}.

\item {\em Random Forest} also from the ``trees'' group. These are 
an ensemble of learning methods for classification, which operate via a multitude of decision trees \citep{bre01}.
 \end{enumerate}

\subsection{Results}
\label{mlr}

We summarise the results in Table \ref{models}. We report on 10-fold cross validation
performance, where the data is split into 10 sets, each
of which will contain $(55 + 43)/10\approx 10$ spectra. Nine of the
datasets are used to train the model, with the resulting classifier
used to test the one remaining dataset. This process is randomised
and repeated ten times with the mean accuracy being reported. 10-fold 
cross validation is typical practice for small data sets, where there
is not enough data to split for training and testing.
\begin{table*} 
\centering 
\begin{minipage}{170mm}
  \caption{The models, without the $z_{\rm weight}$ feature, and their performance for the whole sample and excluding
    the optically thick absorbers.  $P = t_{\rm p}/(t_{\rm p} + f_{\rm p})$ is the precision (the positive predictive
    value), where $t_{\rm p}$ in the number of true positives and $f_{\rm p}$ the number of false positives. $R = t_{\rm
      p}/(t_{\rm p} + f_{\rm n})$ is the recall (the fraction of relevant instances that are retrieved), where $ f_{\rm
      n}$ is the number of false negatives and the F-measure, $F = 2PR/(P+R)$, is the harmonic mean of precision and
    recall. The accuracy, $A$, is the fraction of correctly classified instances.  The mean absolute error,
    $|\overline{\sigma}|$, is a measure of how close the predictions are to the eventual outcomes. The
    $\kappa$-statistic is the chance-corrected measure of agreement between the classifications and the true classes --
    $\kappa > 0$ signifies that the classifier is doing better than predicting by chance and $\kappa =1$ signifies
    completely accurate prediction. }
\begin{tabular}{@{}l   cccccc  cccccc @{}} 
\hline
Algorithm &   $P$\,[\%] & $R$\,[\%]  & $F$\,[\%]  & $A$ \,[\%]  & $|\overline{\sigma}|$  & $\kappa$ & $P$\,[\%] & $R$\,[\%]  & $F$\,[\%]  & $ A$ \,[\%]  & $|\overline{\sigma}|$  & $\kappa$ \\
  & \multicolumn{6}{c}{Whole (55 associated/43  intervening)}    &  \multicolumn{6}{c}{Excluding $\tau_{\rm peak}\geq0.3$ (52 associated/39 intervening)} \\
\hline
Bayesian Network        &81.2  & 80.6  & 80.7 & 80.6 & 0.194 & 0.611 & 83.4 & 83.3 & 83.4 & 83.3 & 0.182 & 0.660\\
Sequential Minimal Optimisation &  80.6 & 78.6 & 78.6 & 78.6 & 0.214 & 0.577 & 78.5& 76.7 & 76.8& 76.97& 0.233 & 0.537\\
Classification Via Regression   & 80.0  & 79.6  & 79.7 & 79.6 & 0.302 & 0.590 & 75.6 & 75.6 & 75.6 & 76.9 & 0.333 & 0.499\\
Logistic Model Tree                 & 80.9 & 80.6 & 80.7 & 80.6& 0.346 & 0.610& 78.3 & 77.8& 77.9 & 77.8 & 0.353 & 0.551 \\
Random Forest                     & 81.6 & 81.6 & 81.6 & 81.6 & 0.305 & 0.422&81.1 & 81.1 & 81.1 & 81.1 & 0.312 & 0.612 \\
\hline
\label{models}
\end{tabular}
\end{minipage}
\end{table*} 
 
Table \ref{confusion}, shows the confusion matrices associated with the models.
\begin{table} 
\caption{The confusion matrices for the models in Table \ref{models}.}
\begin{tabular}{l c c  r r}
\hline
  & \multicolumn{2}{c}{Whole} & \multicolumn{2}{c}{Excluding} \\
  & \multicolumn{2}{c}{sample} & \multicolumn{2}{c}{$\tau_{\rm peak}\geq0.3$} \\
 \hline
Bayesian Network        &
& $\begin{bmatrix}  43 & 12\\ 7 & 36 \end{bmatrix}$ &  
$\begin{bmatrix} 44 & 8 \\ 7& 31 \end{bmatrix}$ \\
 Sequential Minimal Optimisation   &
& $\begin{bmatrix}   39 & 16 \\ 5 & 38 \end{bmatrix}$ &
$\begin{bmatrix} 37 & 15 \\ 6 & 32 \end{bmatrix}$\\
Classification Via Regression   &
& $\begin{bmatrix}   43 & 12 \\ 8 & 35 \end{bmatrix}$ &
$\begin{bmatrix} 41 & 11 \\ 11 & 27 \end{bmatrix}$\\
  Logistic Model Tree                   &
& $\begin{bmatrix}   44 & 11 \\ 8 & 35 \end{bmatrix}$ &
$\begin{bmatrix} 40 & 12 \\ 8 & 30 \end{bmatrix}$\\
Random Forest                         & 
& $\begin{bmatrix}   46 & 9 \\ 9 & 34 \end{bmatrix}$ &
$\begin{bmatrix} 44 & 8 \\ 9 & 29 \end{bmatrix}$\\
\hline
\end{tabular}
\label{confusion}
\end{table} 
These are in the format, 
$\begin{array} {l c c}                                    & \text{\bf Predicted  Ass} & \text{\bf Predicted Int} \\   
\text{\bf Actual Associated} & {\rm TA}  &  {\rm FI}  \\
\text{\bf Actual  Intervening} &  {\rm FA}  &  {\rm TI}  \\
\end{array}$\\
where TA -- true associated, FA -- false associated, TI -- true intervening,  FI -- false intervening.
For example, for the confusion matrix $\begin{bmatrix}  43 & 12\\ 7 & 36 \end{bmatrix}$, 
out of $43 + 12$ associated spectra the model
correctly identifies 43 as associated and 12 erroneously as intervening
and out of $7 + 36$ intervening spectra, 36 are correctly
identified as intervening and 7 erroneously as associated.
The data
in the confusion matrices form the raw data used for reporting
several evaluation measures -- the precision, recall, F-measure and
accuracy (Table \ref{models}). Those in Table \ref{confusion} show that the two classes are
balanced in terms of training and classification. For instance, a
matrix such as $\begin{bmatrix}  90& 0\\ 10 & 0 \end{bmatrix}$, would return a 90\% accuracy, but in this
case the model would have a 100\% recognition rate for one class
and 0\% for the other.

\subsection{Discussion}

Most of the models return a precision, recall, F-measure and accuracy
of $\gapp80$\%, making us optimistic that as more data are collected
by the community, machine learning can provide a useful tool for
classification of redshifted 21-cm absorption spectra.

During our feature selection experimentation we found that
the accuracy of prediction is dominated by the profile width -- models
trained on the FWZI alone return accuracies just a few per-cent
below those reported in Table \ref{models}. Models trained on the three
FWHM features give accuracies $\approx5$\% lower. The removal of all
line-width information features, reduces the accuracy to $\approx50$\%,
i.e. what we would obtain from chance. The number of components,
$n_{\rm g}$, contributes relatively little to the prediction, which is
surprising since we expect the associated profiles to be more complex.
However, the combination of all features do provide the better
performing models, in agreement with the attribute evaluator analysis
(Table \ref{att}).

Another feature we considered is $z_{\rm weight}$. Some experimentation
with classifiers trained using the $z_{\rm weight}$ as an additional feature
improved the precision in some cases (e.g. up to 5.2\%, from
78.5\% to 83.7\% for the Sequential Minimal Optimisation algorithm,
excluding the $\tau_{\rm peak} \geq0.3$ set), while being detrimental in others
(e.g. Bayesian Network). However, since the purpose of this work
is to classify the absorber type without the use of an optical redshift,
the pre-selection of which may introduce a bias (Sect.~\ref{intro}), we
did not consider $z_{\rm weight}$  in the final models (Table~\ref{models}).

The machine learning results are very encouraging and, as
more data are added, we expect the predictive power of such classifiers
to improve. This will prove invaluable as the number of new
21-cm detections becomes too large to feasibly follow-up with optical
observations.

\section{Conclusions}

Forthcoming spectral lines surveys with the next generation of large radio telescopes are expected to detect large
numbers of new redshifted \HI\ 21-cm absorbers. Although the measured redshifts of the absorbing galaxies will be
extremely accurate, due to the phase-locking of radio receivers, without follow-up optical band observations it is
generally not possible to determine whether the absorption arises within the source host or from a galaxy intervening
the sight-line to a more distant radio source. Given these large numbers, and the possibility that optical pre-selection
biases against the detection of cool, neutral gas, it would be of great value to be able to determine the nature of the
absorber based upon the radio data alone.

To this end, we have compiled and digitised the known $z\geq0.1$ \HI\ 21-cm absorbers, converted these to consistent
dimensions (optical depth and velocity) and re-sampled to a common spectral resolution. However, the normalised spectra
in each of the associated and intervening classes exhibit a wide range of profile shapes, not making it possible to
manually ascertain a typical profile shape.  By applying machine learning algorithms, we find that, even for our limited
sample of $\lapp50$ of each type of absorber, the type can be predicted with $\gapp80\%$ accuracy.  As new detections
are made, follow-up optical-band observations will allow us to improve the classifier in preparation for forthcoming
\HI\ 21-cm surveys with the SKA and its precursors.

Although machine learning was invoked to classify the individual spectra, by averaging all those in each class in order to examine the bulk differences, we find:
\begin{enumerate}
 \item That the mean associated profile is wider than the mean intervening profile, with a Kolmogorov-Smirnov
   of the individual widths giving a $4\times10^{-10}$ probability that the associated and intervening velocity distributions are drawn
from the same population. This is consistent with the profile width being the one single feature which lowers the predictive power
of the classifier to that of chance when removed.
\item From a simple model of the \HI\ column density distribution, that the high velocity wings often observed in 
  associated absorption, arise from the sub-pc gas, which appears to be absent in the intervening absorbers.
  This supports the widely proposed  conjecture that the additional component of the associated absorption is due to 
  the dense circumnuclear torus, invoked by unified schemes of AGN. This is also consistent with hypothesis that  associated absorption
arises in AGN (radio galaxies and quasars), where significant amounts of gas are accreted onto the central super-massive black hole, whereas
intervening absorption arises in more quiescent galaxies.
\item The consistency in the mean intervening profile widths to either side of $z\sim1$ (where the sample is split in half),
indicates no kinematical or thermal evolution with redshift. While the \HI\ column density  model is consistent with 
 the bulk of profile broadening being kinematical, rather than thermal, in nature, this result does not support the proposed
increase in the spin temperature of the gas with redshift in damped Lyman-\AL\ absorbers.
\end{enumerate}

\section*{Acknowledgements}

We wish to thank the anonymous referee for their detailed comments and Nathan Holmberg for his advice.
This research has made use of the NASA/IPAC Extragalactic Database (NED) which is operated by the Jet Propulsion Laboratory, California
Institute of Technology, under contract with the National Aeronautics and Space Administration and NASA's Astrophysics
Data System Bibliographic Service. This research has also made use of NASA's Astrophysics Data System Bibliographic Service.


\label{lastpage}

\end{document}